\documentclass[a4paper,11pt]{article} % also: JHEP and JHEP3 %,notoc no-table-of-contents
\pdfoutput=1
\usepackage{subfig}
\usepackage{jheppub} % for details on the use of the package, please
\usepackage[T1]{fontenc} % if needed
\usepackage{microtype}
\usepackage{lmodern}
\usepackage[T1]{fontenc}
\usepackage{amssymb}
\usepackage{mathptmx}
\usepackage{graphicx}
\usepackage{color}
\let\normalcolor\relax % this is the important addition for pdflatex.
\definecolor{myred}{rgb}{0.6,0,0} %usage:  {\textcolor{myred}{Hello World}}
\definecolor{myblue}{rgb}{0,0.2,0.4}
\definecolor{mygreen}{rgb}{0,0.9,0.1}
\definecolor{hc}{rgb}{.9,0.1,0.7}
\definecolor{hcout}{rgb}{.9,0.7,0.9}
\definecolor{Orange}{rgb}{1.,0.65,0.}
\usepackage{amsmath}
\usepackage{rotating}
\usepackage{textcomp}
\usepackage{timestamp} % TR: prints the date and time of latex running

\numberwithin{equation}{section}
\numberwithin{figure}{section}
\numberwithin{table}{section}

\newcommand{\be}{\begin{equation}}
\newcommand{\ee}{\end{equation}}
\newcommand{\bea}{\begin{eqnarray}}
\newcommand{\eea}{\end{eqnarray}}

\title{\boldmath
Simulation of  electron-positron annihilation into hadrons with the event
generator PHOKHARA.
}

\author[a]{H.~Czy\.z~}
\author[a]{M.~Gunia}
\author[b]{J.~H.~K\"uhn}

\affiliation[a]{ Department of Field Theory and Particle Physics,
      Institute of Physics, \\
      University of Silesia, Uniwersytecka 4, PL-40-007 Katowice,
      Poland}
\affiliation[b]{~Institut f\"ur Theoretische Teilchenphysik, \\
   Karlsruher Institut f\"ur Technologie (KIT), D-76128 Karlsruhe, Germany}

\emailAdd{czyz@us.edu.pl}
\emailAdd{guniamichal@gmail.com}
\emailAdd{Johann.Kuehn@KIT.edu}

\abstract{
 The precise determination of the cross section for electron-positron 
annihilation into hadrons is one of the central tasks of ongoing
experiments at low energy colliders. These measurements have to be
complemented by Monte Carlo generators which simulate a large number of
final states and include higher order radiative corrections. With this
motivation in mind the generator PHOKHARA is extended to version 8.0,
 thus allowing for the
simulation of final states with zero, one or two real photons. At the
same time corrections from the emission of one or two virtual photons
are included, such that a full next-to-next-to leading order generator
is available. The stability and consistency of the program is tested.
The results (for muon-pair final states) are compared to the programs
KKMC and MCGPJ and implications for the analysis of various hadronic
final states are investigated.
\color{black}
}

 %, QED, Feynman Integrals}

\preprint{TTP13-020
}
\begin{document}
\maketitle
\flushbottom
% \allowdisplaybreaks

%\tableoffigures

\section{Introduction\label{sec-intro}}
%\addcontentsline{toc}{section}{Introduction}
The importance of a precise measurement of the total cross section for
electron-positron annihilation into hadrons has been emphasised on many
occasions (For recent reviews see e.g. \cite{Actis:2010gg,Hagiwara:2011af,Davier:2010nc,Jegerlehner:2009ry,Harlander:2002ur}).
 Its low energy behaviour 
governs the running of the electromagnetic coupling from the Thompson
limit to higher energies and is, therefore, a decisive input for all 
precision analyses of electroweak interactions. Indeed, it may well be
one of the limiting factors for the interpretation of the Standard
Model, in particular the test of the relation between masses of the 
top quark, the W and the Higgs boson \cite{ALEPH:2010aa,Eberhardt:2012gv}. 
It is, furthermore, one of
the limiting ingredients for the theory prediction of the anomalous
magnetic moment of the muon. Last not least, this cross section gives
crucial input to dedicated analyses based on perturbative QCD (pQCD),
 be it the determination of 
the strong coupling constant $\alpha_s$ (for a review see 
\cite{Chetyrkin:1994js,Chetyrkin:1996ia}), quark mass determinations
\cite{Chetyrkin:2009fv} or low energy quantities like the pion or nucleon form factors.
At high and intermediate energies a purely perturbative treatment of the
total cross section is assumed to provide sufficiently precise
predictions, however at low energies and in the charm- and
bottom-quark threshold region  no ab-initio prediction based on perturbative
QCD (pQCD) can be made.

To determine the total cross section, one may either perform an
inclusive measurement, or identify individually all the different
multi-hadron final states. From the theoretical  side the inclusive
cross section can be well predicted on the basis of perturbative QCD (at
least for properly chosen energy regions), exclusive channels carry
more detailed information on form factors, resonances, isospin symmetry
and breaking. In view of the importance of these measurements at low
energies, where the number of different exclusive modes is still
limited, and considering the fact, that this region plays the dominant
role for the aforementioned theoretical investigations, it seems useful
to develop a Monte Carlo generator which is tailored to the simulation
of exclusive hadronic final states.

The measurement of the annihilation cross section proceeds in two
conceptually different ways. The traditional and most obvious method
is based on a variable center of mass energy of the electron-positron
collider which allows the experiment to scan through an
energy range which is trivially dictated by the available beam 
energy (scanning mode). 
As a second alternative one may use the
''radiative return'', exploiting the fact that initial state radiation
leads to a variable ''effective'' energy and invariant mass of the
hadronic system. In view of the large luminosity of current
electron-positron colliders, the loss in cross section, resulting from
the factor of $\alpha$ to be paid for the photon emission, is compensated
by the advantage of running at a stable fixed beam energy.

Various generators have been developed to simulate events with radiative return
into a variety of hadronic and leptonic final states in leading order
(LO) \cite{Binner:1999bt,Czyz:2000wh} and NLO 
\cite{Kuhn:2002xg,Rodrigo:2001kf,Czyz:2002np}.
 Also for the scanning mode
a number of precision generators have been developed
which include electromagnetic corrections from initial state radiation
in next-to-leading (NLO) and partly even next-to-next-to-leading (NNLO) order
 \cite{Jadach:1999vf,Arbuzov:1997je,Arbuzov:1997pj,Jadach:2000ir,Arbuzov:2005pt}.
Note, that the counting of orders is different for the scanning mode
and the radiative return, since one photon emission constitutes the
leading order process in the latter, while this same process contributes
to the NLO corrections in the scanning mode. It is the aim of this work
to construct a new NNLO generator for the scanning mode, which is based on the
already existing PHOKHARA 7.0 \footnote{Manuals for the usage of of the 
different
versions of PHOKHARA up to v8.0 can be found under
ific.uv.es/\textasciitilde rodrigo/phokhara/} , as far as radiative corrections from one and two
photon emission are concerned. Evidently, the only missing ingredient,
the two loop virtual corrections, are available in the literature since
long \cite{Berends:1987ab} and can be implemented into our generator in a
straightforward way. 

At present, all generators simulating the scanning mode
have been constructed for leptonic final states only and to a limited extent
for two body ($\pi\pi$ and $KK$) hadronic states. 
In contrast, the Monte Carlo event generator PHOKHARA has been designed
from the beginning to
simulate exclusive hadronic final states, using specific
models for the form factors. Indeed many two-, three- and four-body final states
have been implemented by now
(For the detailed listing of the final states and the underlying
assumptions about the form factors see Section \ref{sec-nnlo}.)
 As said above, it is a
fairly straightforward task to extend PHOKHARA 7.0 such that it is
applicable for the scanning mode, thus providing a generator which
includes the full NNLO corrections, at least as far as initial state
radiation is concerned.

This program will be complementary to the two main generators currently in
use, KKMC 4.13 \cite{Jadach:1999vf,Jadach:2000ir} and MCGPJ \cite{Arbuzov:1997je,Arbuzov:1997pj,Arbuzov:2005pt}.
 The former is based on YFS
exponentiation, thus includes the emission of an arbitrary number of soft
photons and hence a certain class of corrections which may be important
for the measurement of a rapidly varying cross section, in
particular close to a narrow resonance. This feature is  not implemented in
PHOKHARA. On the other hand, in contrast to PHOKHARA 8.0 the currently
available version 4.13 of KKMC does not include the full NNLO corrections in the
region of hard real photon emission (Although these corrections, 
originally evaluated in \cite{Kuhn:2002xg} and recalculated 
 in \cite{Yost:2005ju,Jadach:2006fx}, were
implemented in a private version of the program and
used for analysis in \cite{Jadach:2005gx}.).
 Furthermore, KKMC~4.13 is restricted to leptonic final states and thus
cannot serve for an analysis of the multitude of hadronic states
mentioned above \footnote{A different, unpublished, version of KKMC
 \cite{KKMC4.16}
implements a number of hadronic final states, albeit with less elaborate
hadronic matrix elements and NLO ISR only.}. 

MCGPJ, on the other hand, is built on the usage of structure
function method, an approach which does not
allow to simulate the proper kinematics in the case of hard, non-collinear
photon emission. Furthermore, among the hadronic final states only
$\pi^+\pi^-$, $K^+K^-$ and $\bar K^0 K^0$ are included.

Let us conclude this section with a brief comment concerning final state
radiation. Inclusion of one-photon real and virtual corrections is
straightforward, as far as lepton final states are concerned. For two
body hadronic states an ansatz based on point-like pions, kaons and
protons has been implemented in PHOKHARA from the very beginning. 
This ansatz has been
successfully compared to data \cite{Muller:2009pj} as long as relatively soft
photons of several hundred MeV are concerned. No further comparison
between data and model has been performed to our knowledge. For
multi-hadron final states use of PHOTOS (\cite{Barberio:1993qi}
 and updates)
is recommended.
%==============================================================================
\section{The NNLO corrections\label{sec-nnlo}}
%\addcontentsline{toc}{section}{The NNLO corrections}

  As stated above, we would like to develop an event generator
 for the processes
 $e^+e^- \to  hadrons$ and $e^+e^- \to \mu^+\mu^-$, which relies on
 fixed order formulae and simulates exact kinematics.
 The starting point for this development  is the event generator PHOKHARA
 \cite{Rodrigo:2001kf}, and in fact its latest version PHOKHARA7.0 
 \cite{Czyz:2010hj} has been used for this purpose. The generator profits from 
 the dedicated and continuously  updated hadronic currents and the well tested
 implementation of QED radiative corrections. The master formula for 
 the fixed order calculations, which is now implemented in the code
 reads

   \begin{eqnarray}\label{all-fermions}
    d\sigma (e^+e^- \to {\rm hadrons} + {\rm photons}) &=& 
  d\sigma (e^+e^- \to {\rm hadrons})
  \nonumber \\
 &+& d\sigma (e^+e^- \to {\rm hadrons} + {\rm one\ hard\ photon})
 \nonumber \\
 &+& d\sigma (e^+e^- \to {\rm hadrons} + {\rm two\ hard\ photons}) \ .
   \end{eqnarray} 

The present version 8.0 of the program is limited to initial state emission and only
 the photon emission from electron and positron is taken into account.
Emission of real and virtual lepton pairs will be treated in a later version. 
The radiative corrections in $d\sigma (e^+e^- \to {\rm hadrons} + {\rm one\ hard\ photon})$ are described in \cite{Rodrigo:2001jr,Kuhn:2002xg} 
and the real emission of two photons in \cite{Rodrigo:2001kf}. 
 The  implementations of various
 hadronic modes with up-to-date hadronic currents modelling are discussed in
 \cite{Czyz:2010hj} for $\pi^+\pi^-$, $K^+K^-$ and $\bar K^0 K^0$  final states,
in \cite{Czyz:2004ua} for $\bar p p$ and $\bar n n$ final states, 
 in \cite{Czyz:2005as} for $\pi^+\pi^-\pi^0$ final state, in \cite{Czyz:2008kw}
 for $\pi^+\pi^-2\pi^0$ and $2\pi^+2\pi^-$ final states and 
 in \cite{Czyz:2007wi} for $\Lambda(\to\pi^-p)\bar \Lambda(\to \pi^+\bar p)$ 
 final state. Since the implementation of the $\eta \pi^+\pi^-$ channel 
 was never
  documented in the literature we add a description of this hadronic current 
 in  Appendix (\ref{app-eta}).
For the readers convenience the most important formulae  are repeated in
the following. 

\vskip 0.5 cm
 {\it i) single photon emission in LO and NLO:}
\vskip 0.5 cm
 
 The differential rate in $d\sigma (e^+e^- \to {\rm hadrons} + {\rm one\ hard\ photon})$
is \cite{Rodrigo:2001kf}
cast into the product of a leptonic and a hadronic tensor and the 
corresponding factorised phase space:
\begin{equation}
d\sigma(e^+e^- \to {\rm hadrons} + {\rm one\ hard\ photon}) = \frac{1}{2s} L_{\mu \nu} H^{\mu \nu}
d \Phi_2(p_1,p_2;Q,k_1) d \Phi_n(Q;q_1,\cdot,q_n) 
\frac{dQ^2}{2\pi}~,
\label{1ph}
\end{equation}
where $d \Phi_n(Q;q_1,\cdot,q_n)$ denotes the hadronic 
$n$-body phase space including all statistical factors 
and $Q^2$ is the invariant mass of the hadronic system.

The physics of the hadronic system, whose description is 
model-dependent, enters only through the hadronic tensor 
\begin{equation}
H^{\mu \nu} = J^{\mu} J^{\nu +}~,
\end{equation}
where the hadronic current has to be parametrised through form factors,
  which depend on the four momenta of the final state hadrons. The form of the 
 leptonic tensor, including the one-loop correction and emission
 of a second soft photon as well as the vacuum polarisation corrections,
  can be found in \cite{Rodrigo:2001kf} (Eq.(5)). 

 \vskip 0.5 cm
{\it ii) two photon emission:}
\vskip 0.5 cm

  For the evaluation of $d\sigma (e^+e^- \to {\rm hadrons} + {\rm two\ hard\ photons})$ the helicity
 amplitude method was used as described in Section 3 of \cite{Rodrigo:2001kf}.
 This part contains as well  the vacuum polarisation corrections.

\vskip 0.5 cm
{\it iii) zero photon emission in LO, NLO and NNLO:}
\vskip 0.5 cm

 The new part, $d\sigma (e^+e^- \to {\rm hadrons})$, added to the code is
 written as 

\begin{equation}
d\sigma(e^+e^- \to {\rm hadrons}) = \frac{1}{2s} L^0_{\mu \nu} H^{\mu \nu}
 d \Phi_n(p_1+p_2;q_1,\cdot,q_n) ~,
\end{equation}
with the same hadronic tensor, but calculated at a different from Eq.(\ref{1ph})
 kinematic point. The leptonic tensor contains the virtual and soft radiative
 corrections up to the second order \cite{Berends:1987ab}

\begin{eqnarray}
L_{\mu \nu}^{0} = 4( p_{1 \mu}p_{2 \nu} -g_{\mu \nu} \frac{s}{2} +   p_{1 \mu}p_{2 \nu} ) 4 \pi \alpha  |\frac{1}{1-\Delta_{VP}(s)}|^2 (1+ \Delta )
\end{eqnarray} 
where $\Delta_{VP}(s)$ is the vacuum polarisation correction,

\begin{eqnarray}
\Delta = \Delta_{virt,1ph} + \Delta_{soft,1ph} + \Delta_{virt,2ph} + \Delta_{soft,2ph} + \Delta_{virt,soft,1ph}
\end{eqnarray}

\begin{eqnarray}
\Delta_{soft,1ph} =\frac{\alpha}{\pi}(
  \frac{1}{2} \log^2{(s/m_e^2)} + 2 \log{(2 w)} ( \log{(s/m_e^2)} -1) -2 \zeta _2
  )
\end{eqnarray}
with $w = E_\gamma^{min}/\sqrt{s}$;

\begin{eqnarray}
\Delta_{virt,1ph} =2 {\rm Re} (F_1)
= \frac{2\alpha}{\pi} (-\log^2{(s/m_e^2)}/4 +3\log{(s/m_e^2)} -1 +2\zeta _2  )
\end{eqnarray}
\begin{eqnarray}
{\rm Im (F_1)} = \alpha (\log^2{(s/m_e^2)}/2 - \frac{3}{4}  )
\end{eqnarray}
\begin{eqnarray}
\Delta_{soft,2ph} =\frac{\Delta_{soft,1ph} ^2}{2}
\end{eqnarray}

\begin{eqnarray}
\Delta_{virt,soft,1ph} =\Delta_{soft,1ph} \Delta_{virt,1ph}
\end{eqnarray}

\begin{eqnarray}
\Delta_{virt,2ph} =|F_1|^2 + 2 {\rm Re} (F_2)
\end{eqnarray} where:

\begin{eqnarray}
{\rm Re} (F_2) =(\frac{\alpha}{\pi})^2 (\log^4{(s/m_e^2)} - 3\frac{\log^3{(s/m_e^2)}}{16} + \log^2{(s/m_e^2)}(\frac{17}{32} - \frac{5 \zeta _2}{4}) \\ \nonumber
 + \log{(s/m_e^2)}(-\frac{21}{32} + 3 \zeta _2 + \frac{3 \zeta _3}{2}) - 3 \zeta_2 \log{(2)} - \frac{\zeta_2}{2} + \frac{405}{216})
\end{eqnarray}

\begin{figure}[ht]
\begin{center}
\vspace*{-5.0cm}
\includegraphics[width=0.6\textwidth]{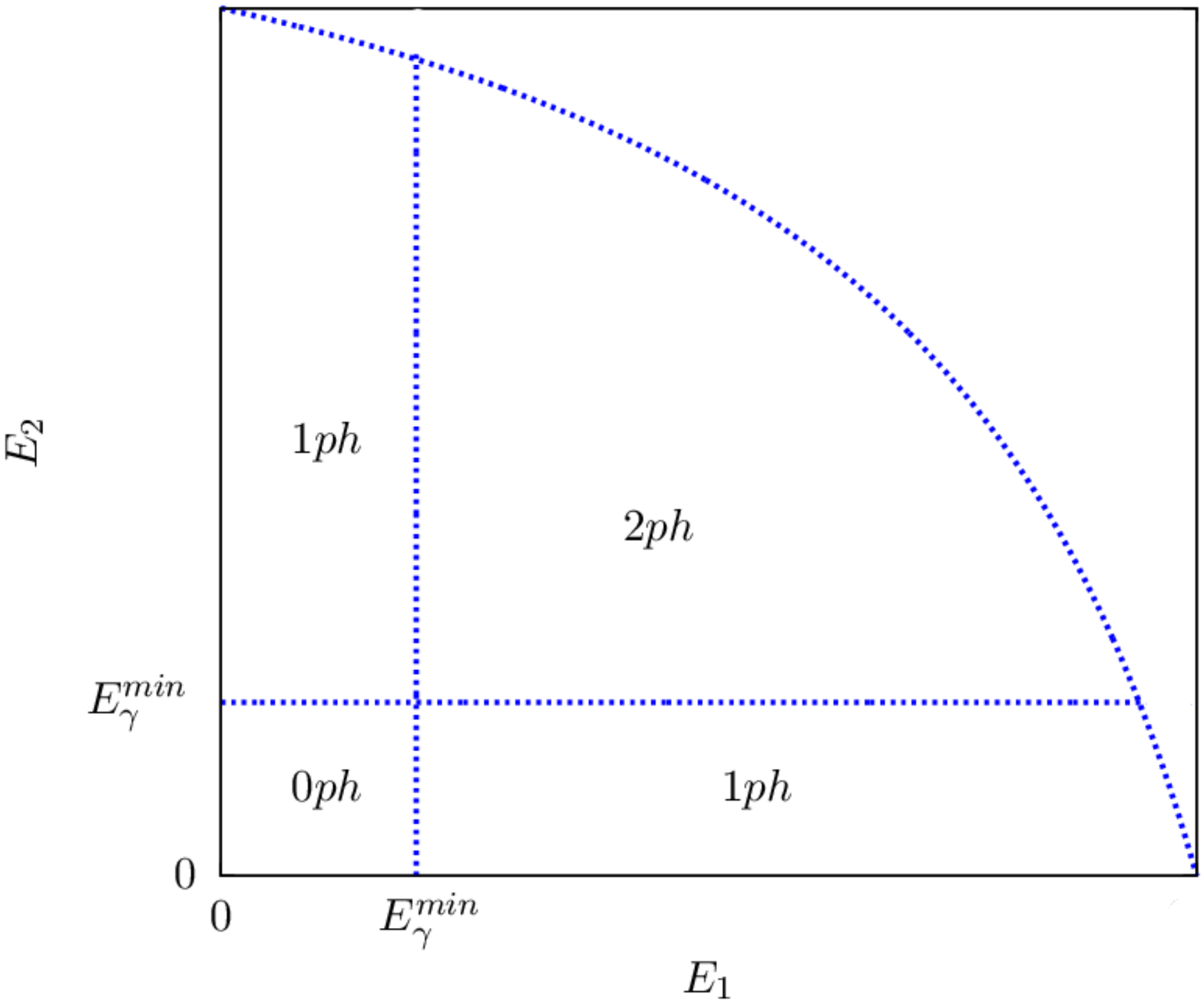}
\end{center}
%\vspace*{-1.0cm}
\caption[]{Division of contributions from two photon phase space into three
 parts of the cross section. $0ph$-part is added into $d\sigma (e^+e^- \to {\rm hadrons})$; $1ph$-part is added into $d\sigma (e^+e^- \to {\rm hadrons}+ {\rm one\ hard\ photon})$; $2ph$-part is calculated in $d\sigma (e^+e^- \to {\rm hadrons}+ {\rm two\ hard\ photons})$
  }
\label{phspplot}
\end{figure}

  We use a bit different from \cite{Berends:1987ab} 
  division of the two-photon phase space (see Fig. \ref{phspplot})
  into soft-soft, soft-hard and hard-hard parts. Hence the soft photon 
  contribution  coming from two soft photons and calculated analytically
  is also different. This division is more suitable for the implementation 
  into our generator as we generate directly the photon energies.

 To generate the phase space of the hadrons~+~two photons,
 the program generates first the invariant mass of the hadronic system $q^2$
 and the angles of the photons in the center-of-mass-frame of
  the initial fermions.
 Then the energy of one of the photons is generated in this frame and
 the second is calculated from the relation 

 \begin{equation}
  q^2 = s-2(E_1+E_2)\sqrt{s} + 2 E_1 E_2 (1- \cos\theta_{12}),
\label{pppp}
 \end{equation}
 where
 $E_1,E_2$ are the energies of the photons and $\theta_{12}$ is the angle
 between their momenta. The curve from Eq. (\ref{pppp}) gives the boundary of
 the allowed energies 
  in Fig. \ref{phspplot}.

\section{The generator and its tests\label{sec-virt-nnlo}}

%\addcontentsline{toc}{section}{The generator and its tests}

  The tests presented here concentrate on the testing of the implementation
 of the new (zero-photon emission)  part since the other parts were already well
  tested. 

\begin{figure}[ht]
\begin{center}
\vspace*{-4.0cm}
%\vspace*{2.0cm}
%\includegraphics[width=.8\textwidth]{A_IRsoft-2}
\hspace*{-2.0cm}\subfloat{\includegraphics[width=.8\textwidth]{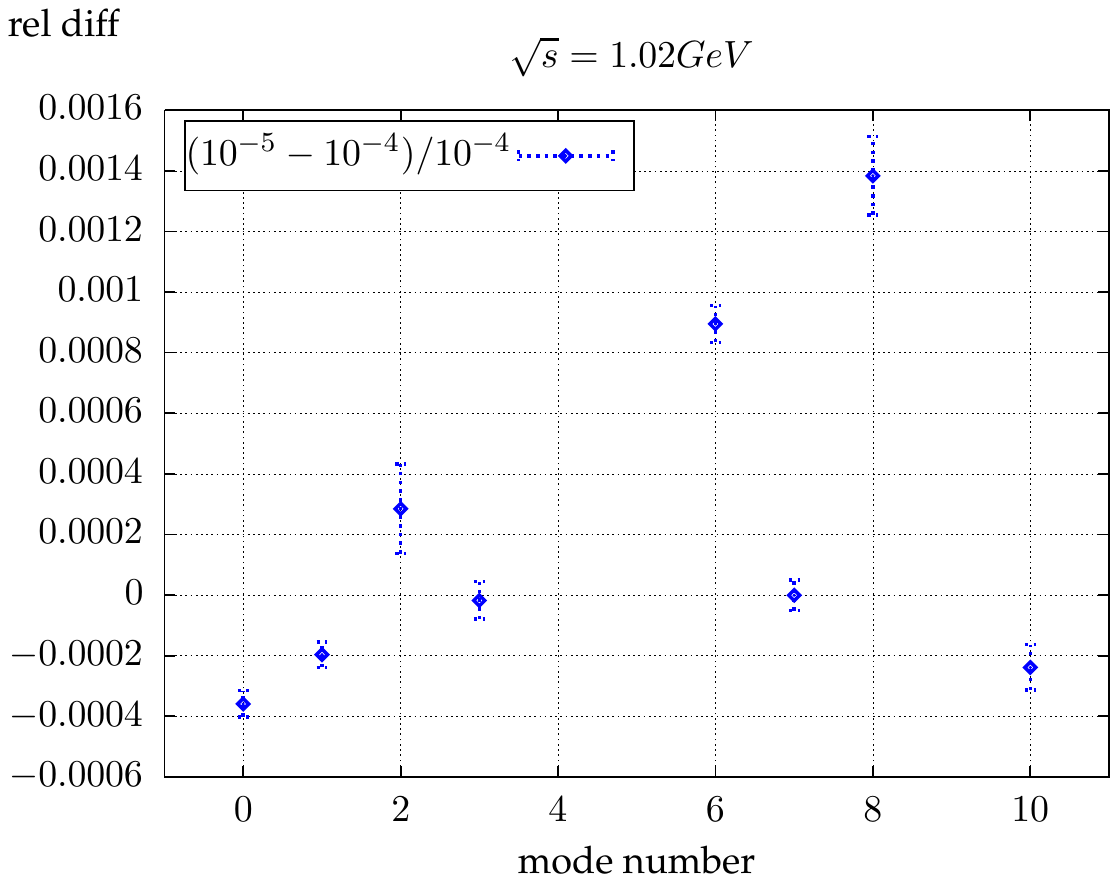}}
\hspace*{-4.0cm}\subfloat{\includegraphics[width=.8\textwidth]{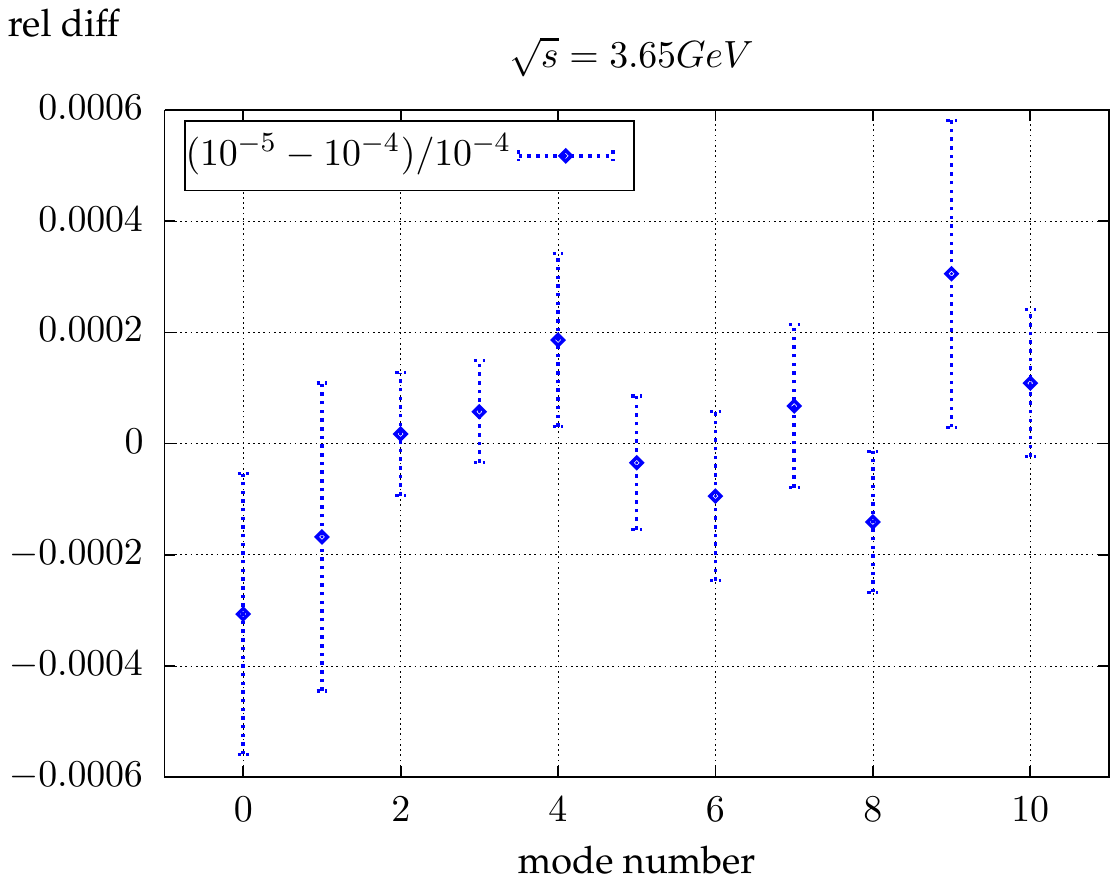}}
\end{center}
\vspace*{-8.0cm}
\caption[]{The difference between integrated cross sections with separation 
parameter $w=10^{-4}$ and $w=10^{-5}$ for 10 modes in PHOKHARA 8.0:
 0 ($\mu^+\mu^-$), 1 ($\pi^+\pi^-$), 2 ($2\pi^0\pi^+\pi^-$), 3 ($2\pi^+2\pi^-$), 4 ($p \bar p$), 5 ($n \bar n$), 6 ($K^+K^-$), 7 ($K^0 \bar K^0$), 8 ($\pi^0\pi^+\pi^-$), 9 ($\Lambda(\to \pi^- p) \bar \Lambda(\to \pi^+\bar p)$), 10 ($\eta\pi^+\pi^-$) }
\label{cutstests}
\end{figure}

For the sum of all contributions we  have tested  
 the independence of the integrated cross section from the separation
  parameter between soft and hard parts. The results are summarised
  in Fig. \ref{cutstests}. The recommended cut to be used is  $w=10^{-4}$,
  while for $w=10^{-5}$ we start to observe negative weights and the
  result is obtained using weighted events.
  Perfect agreement between the results  
  for $\sqrt{s}=3.65 $~GeV (in fact we have performed tests for several energies
 with similar results) demonstrates that the generation in the
 soft photon region is implemented properly. The small disagreement
  at $\sqrt{s}=1.02 $~GeV is due to the presence of a relatively narrow
 resonance $\phi$ and it disappears when one further  decreases  the $w$
  cut. This difference together with the appearing of the negative weights
  also indicates that the missing multi-photon emission is important in this
  region. If one runs with $w=10^{-4}$ the difference to  $w=10^{-5}$ is
  an estimate of additional error of the generator (the error estimate will
  be given at the end of this section).

  Apart of many technical tests of the one- and two-photon parts
  we would like to recall a comparison presented in \cite{Jadach:2005gx}, where
  it was shown that the muon pair  invariant  mass distributions obtained
  with PHOKHARA 2.0 on the one hand and with KKMC upgraded for this purpose 
  on the other hand are in excellent
  agreement. 
This is valid except in the kinematical region
where the muon pair invariant mass is very close to $\sqrt{s}$.
  In this case multi-photon emission
  is most important and
  leads to differences up to a few percent. 

\begin{figure}[ht]
\begin{center}
\vspace*{-4.0cm}
\hspace*{-2.0cm}\subfloat{\includegraphics[width=.8\textwidth]{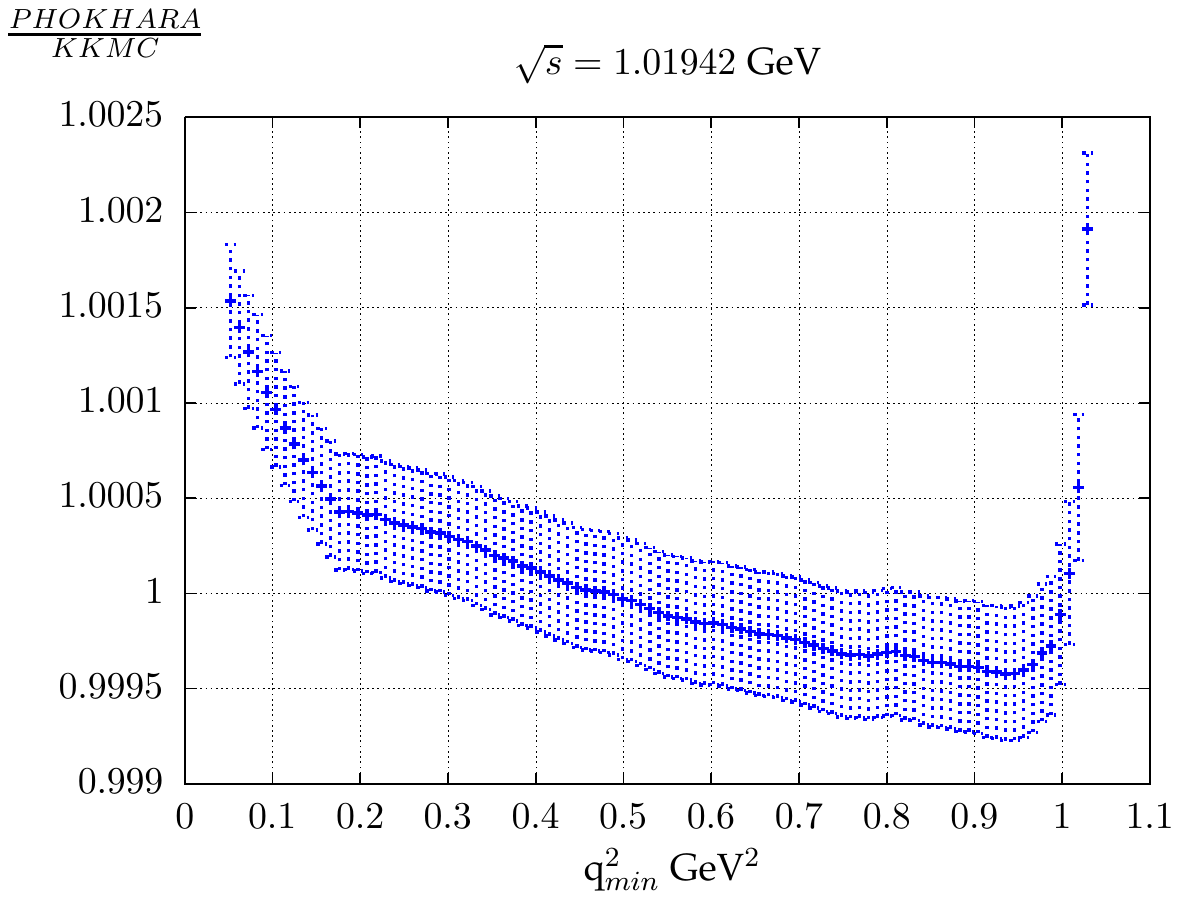}}
\hspace*{-4.0cm}\subfloat{\includegraphics[width=.8\textwidth]{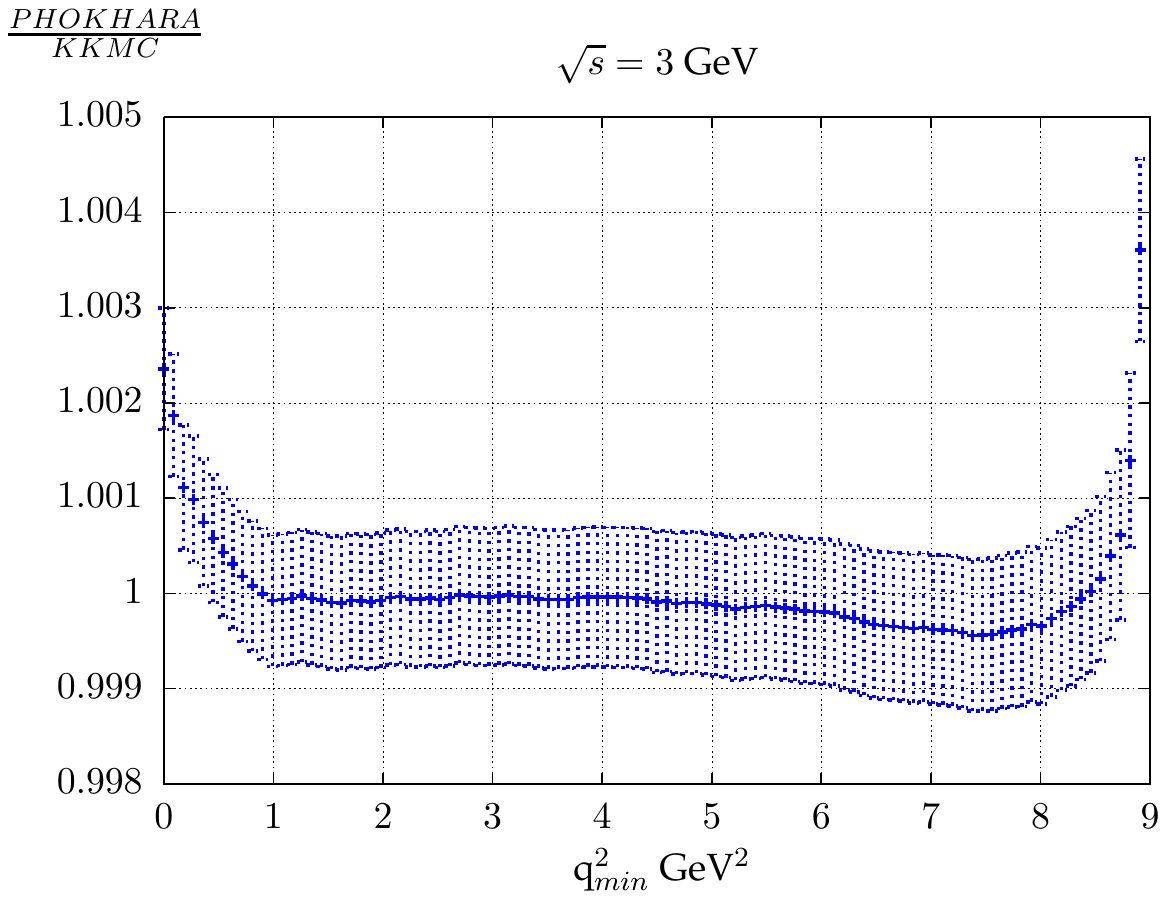}}
\end{center}
\vspace*{-8.0cm}
\caption[]{Ratio of the integrated cross section obtained by KKMC 4.13
and PHOKHARA 8.0
as a function of the lower limit of the muon pair invariant mass $q^2_{min}$. }
\label{intKKPH}
\end{figure}

  For any scan experiment one
  measures the photon-inclusive cross section with a loose cut 
  on the invariant mass of the hadronic/muonic system, say 
  $q^2_{min} > s/2  $ .
  The agreement between the two codes KKMC 4.13 and PHOKHARA 8.0 is even better
  for this type of comparison. This is demonstrated in  Fig. \ref{intKKPH},
  where we show the ratio of the integrated cross sections
   $ \int_{q^2_{min}}^s \frac{d\sigma}{dq^2} dq^2$ as a function of
  $q^2_{min}$ for two values of s. 
   The disagreement
  for very tight selection cut ($q^2_{min}$ close to $s$) is due to missing 
  multi-photon corrections in PHOKHARA 8.0, while for small $q^2_{min}$ the
  missing terms in public version of KKMC 4.13 are responsible 
 for the difference.
 For the realistic choice of the cut $q^2_{min}$ the agreement is better
   than 0.2 permille.

\begin{figure}[ht]
\begin{center}
%\vspace*{-4.0cm}
%\includegraphics[width=.8\textwidth]{A_IRsoft-2}
\subfloat{\includegraphics[width=.5\textwidth]{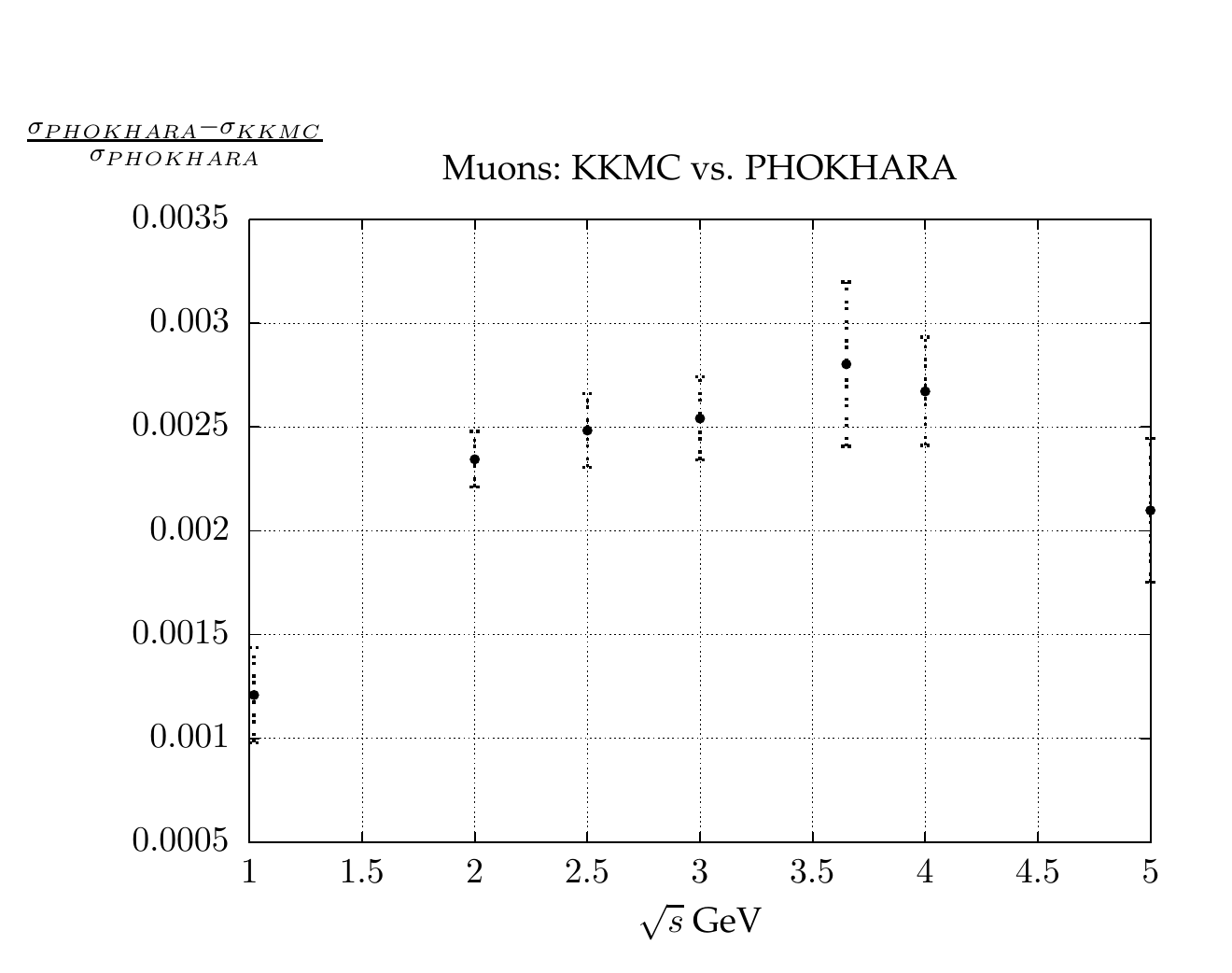}}
\subfloat{\includegraphics[width=.5\textwidth]{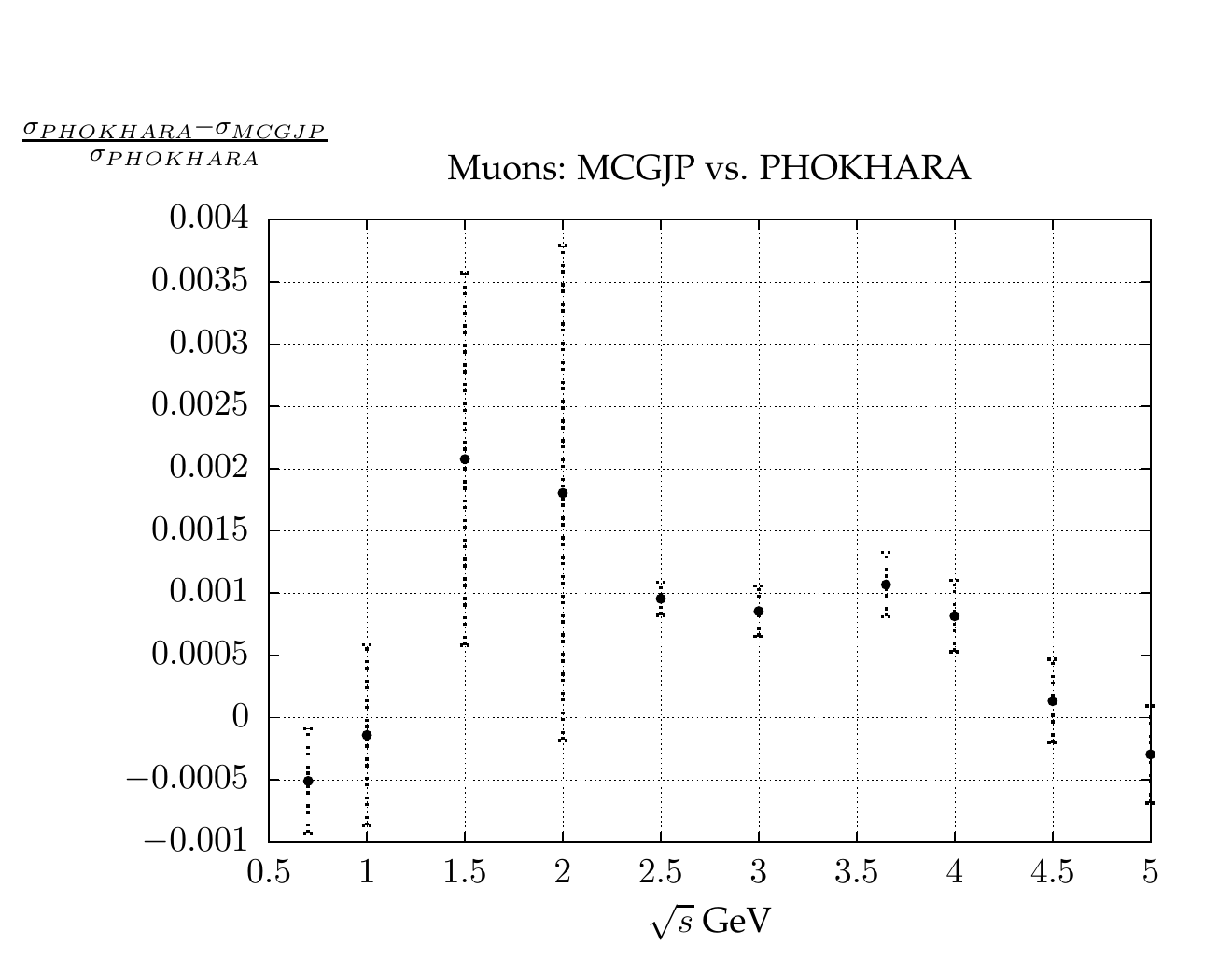}}
\end{center}
%\vspace*{-8.0cm}
\caption[]{ The size of the relative difference between PHOKHARA 8.0, 
 KKMC 4.13 and MCGPJ for muon pair production.}
\label{reldiffall}
\end{figure}

  The difference between the predictions of PHOKHARA 8.0 and KKMC 4.13
  for 
 the inclusive cross sections for muon production
 depends only slightly on the center-of-mass energy as can be observed
 in Fig. \ref{reldiffall}. The difference is even smaller for MCGPJ. However,
 below 2 GeV the version of the MCGPJ code \cite{MCGPJ} we were using
 was not stable numerically when trying to improve the accuracy
 of the result. The indicated errors are  the smallest
 we were able to get. 

\begin{figure}[ht]
\begin{center}
\vspace*{-4.0cm}
\includegraphics[width=0.95\textwidth]{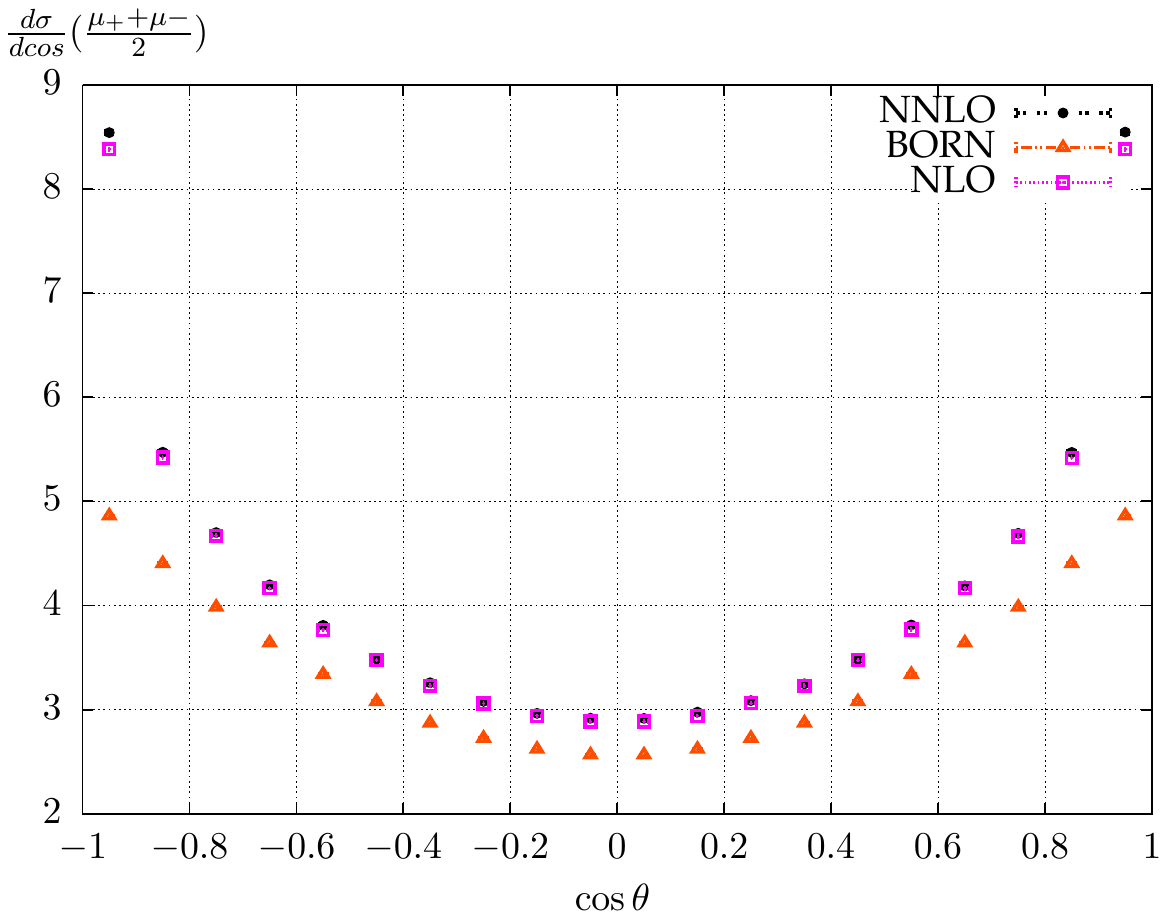}
\end{center}
\vspace*{-10.0cm}
\caption[]{{The size of NLO and NNLO corrections to the muon polar angle
 distribution. }}
\label{muth}
\end{figure}
\begin{figure}[ht]
\begin{center}
\vspace*{-4.0cm}
\hspace*{-2.0cm}\subfloat{\includegraphics[width=.8\textwidth]{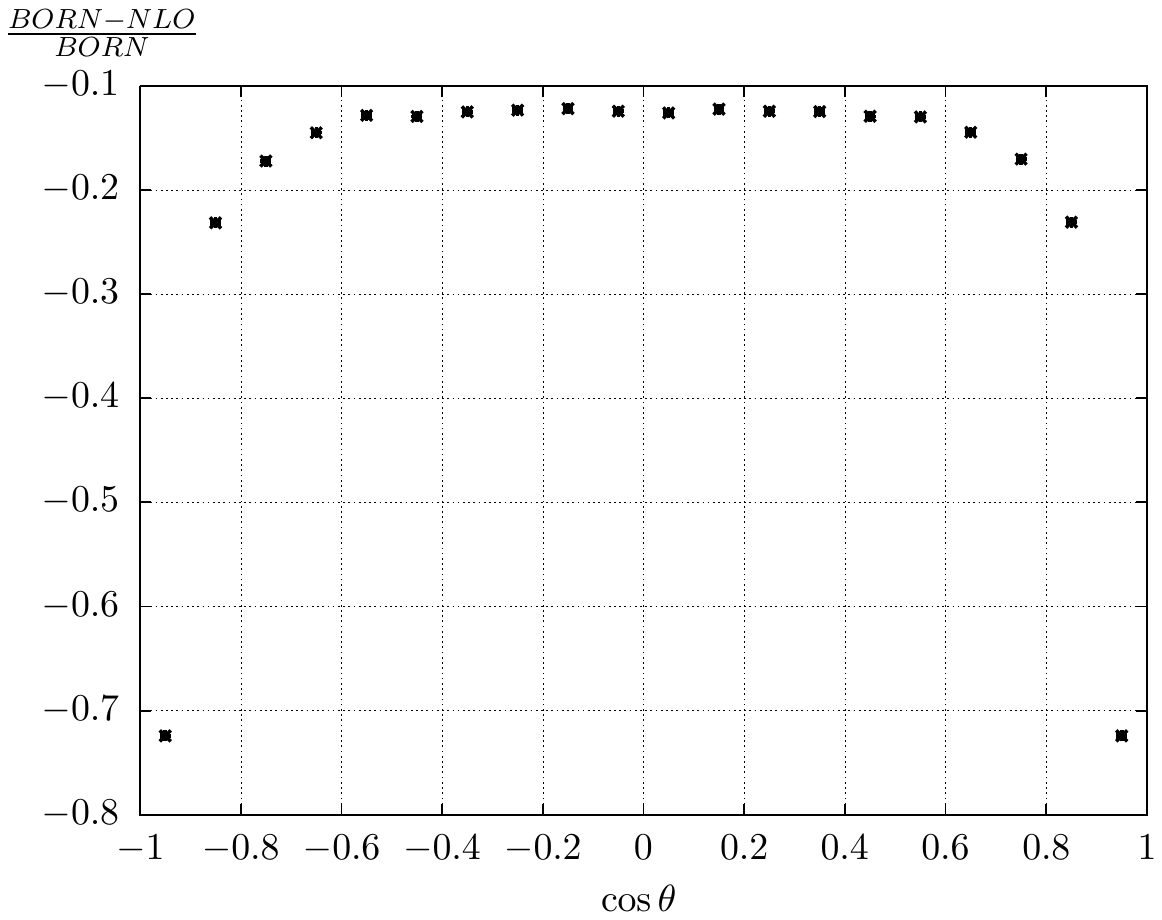}}
\hspace*{-4.0cm}\subfloat{\includegraphics[width=.8\textwidth]{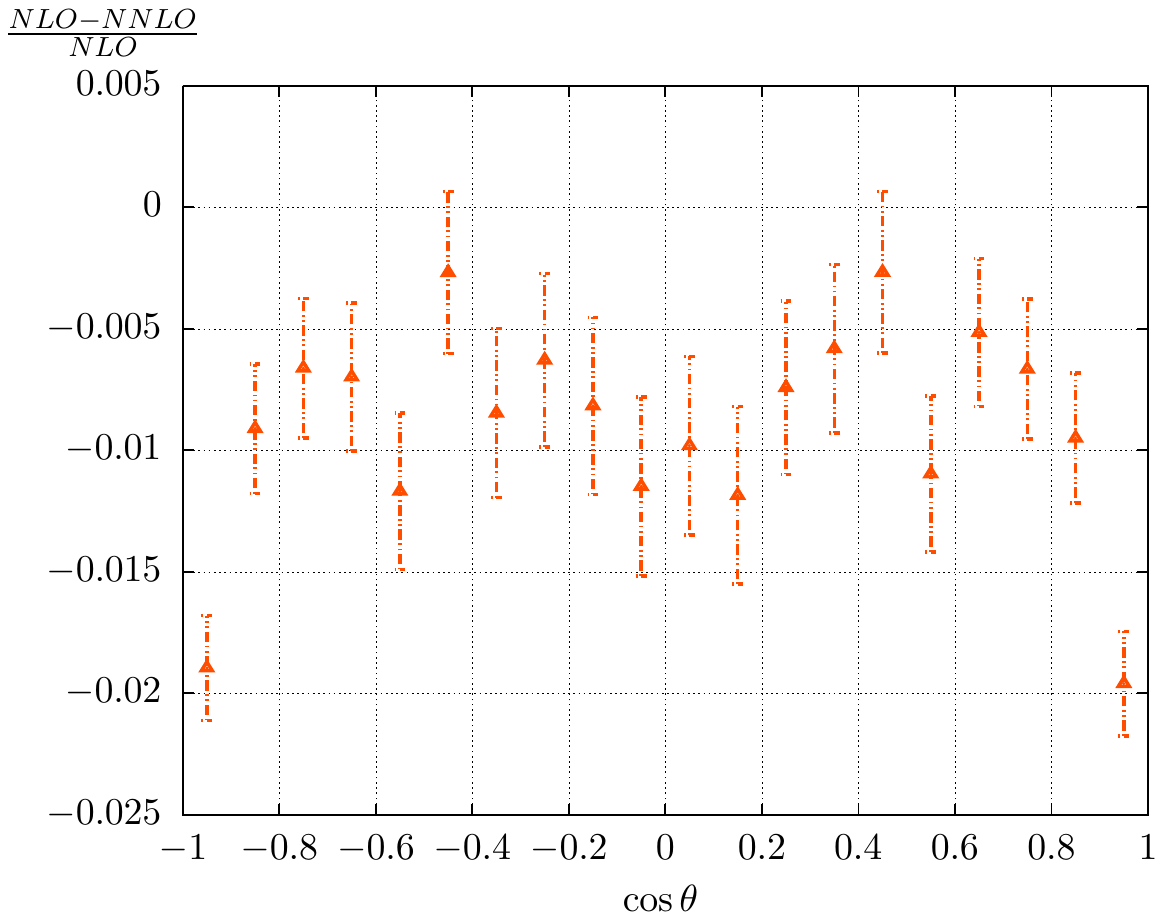}}
\end{center}
\vspace*{-8.0cm}
\caption[]{ The size of NLO and NNLO corrections to the muon polar angle
 distribution.}
\label{relmuth}
\end{figure}

 For many experimental applications of an event generator, 
 for example studies of acceptance corrections, the fully differential
 cross section is needed.
A typical modification of an angular distribution by the radiative 
 corrections is shown for muon polar angle distribution in Fig. \ref{muth}.
 The NNLO corrections, even if they are small (Fig. \ref{relmuth}), are 
  relevant in the era of precision hadronic physics \cite{Actis:2010gg}.
\begin{figure}[ht]
\begin{center}
\vspace*{-4.0cm}
\hspace*{-2.0cm}\subfloat{\includegraphics[width=.8\textwidth]{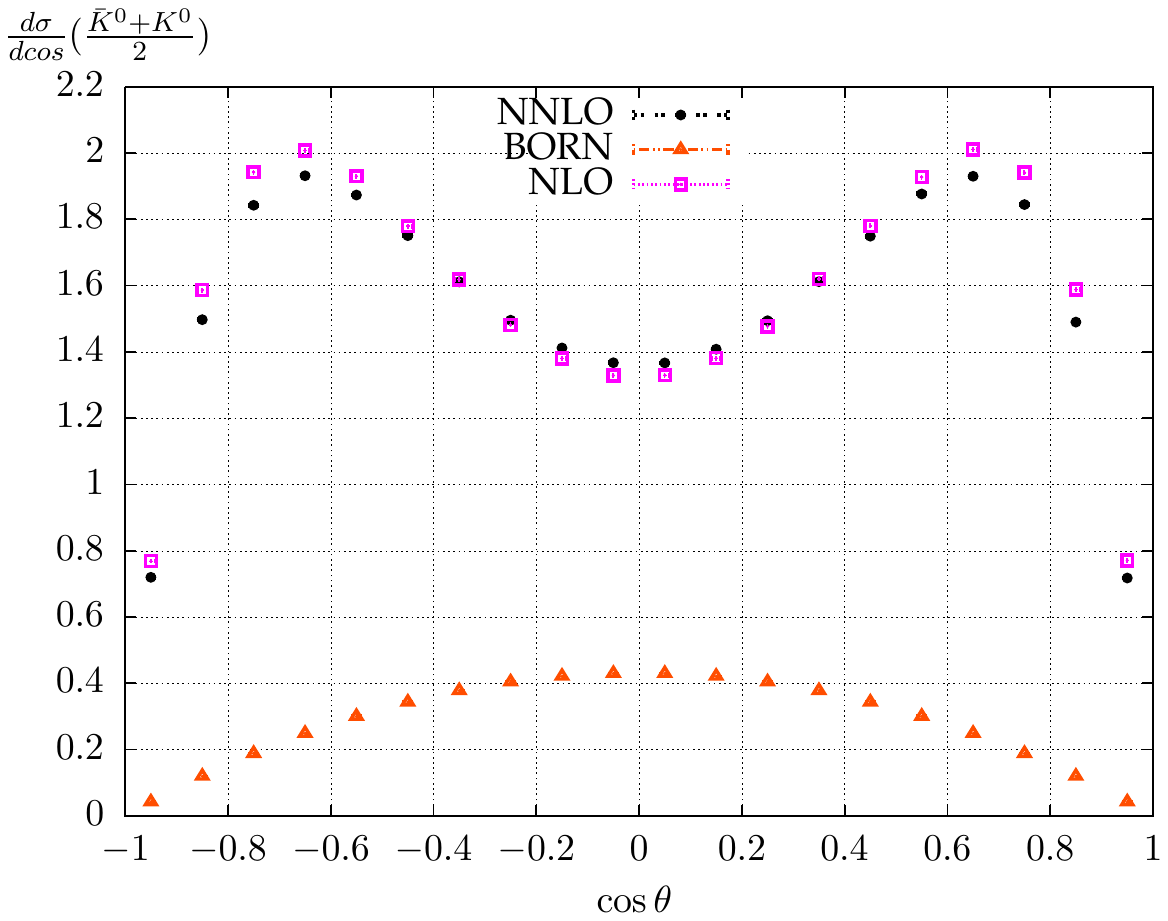}}
\hspace*{-4.0cm}\subfloat{\includegraphics[width=.8\textwidth]{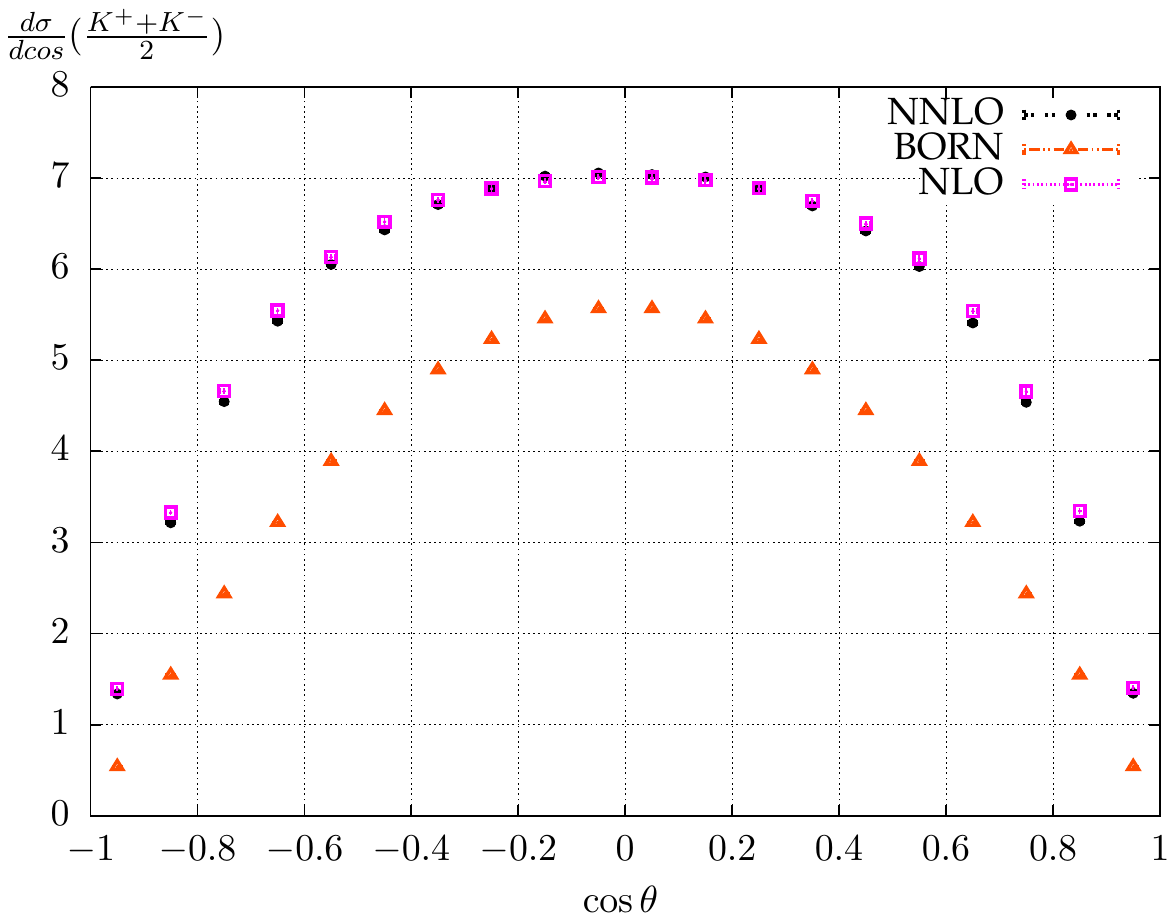}}
\end{center}
\vspace*{-8.0cm}
\caption[]{ The size of NLO and NNLO corrections to the kaon polar angle
 distribution (left plot neutral kaons, right plot charged kaons).}
\label{angkaons}
\end{figure}

 For some particular cases, when for example at a nominal energy
  of the experiments the form factor is small, the radiative corrections
  might be bigger than the LO result. It is shown 
  in Fig. \ref{angkaons} for neutral kaons for the
 center of mass energy 1.2 GeV. 
The large correction reflects the large number of events from the radiative
 return to the $\Phi$ with subsequent decay into neutral kaons, i.e.
 from $e^+e^-\to\gamma\Phi(\to K^0 \bar K^0)$. For comparison we also show
 the corresponding plot for charged kaons. The charged kaon form factor at
1.2~GeV is significantly larger than the one for the neutral kaon and 
the relative impact of $\gamma\Phi(\to K^+  K^-)$ is correspondingly smaller.

\begin{figure}[ht]
\begin{center}
\vspace*{-4.0cm}
\hspace*{-2.0cm}\subfloat{\includegraphics[width=.8\textwidth]{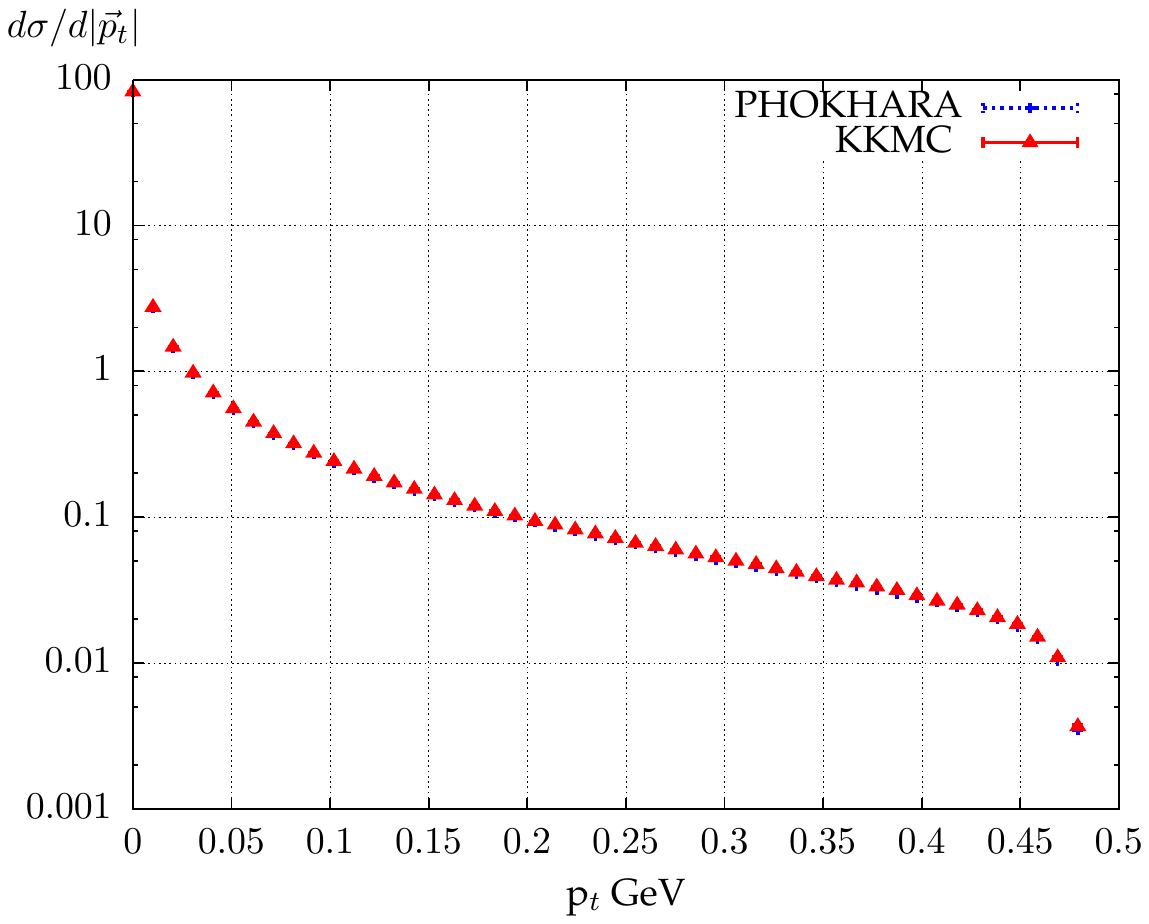}}
\hspace*{-4.0cm}\subfloat{\includegraphics[width=.8\textwidth]{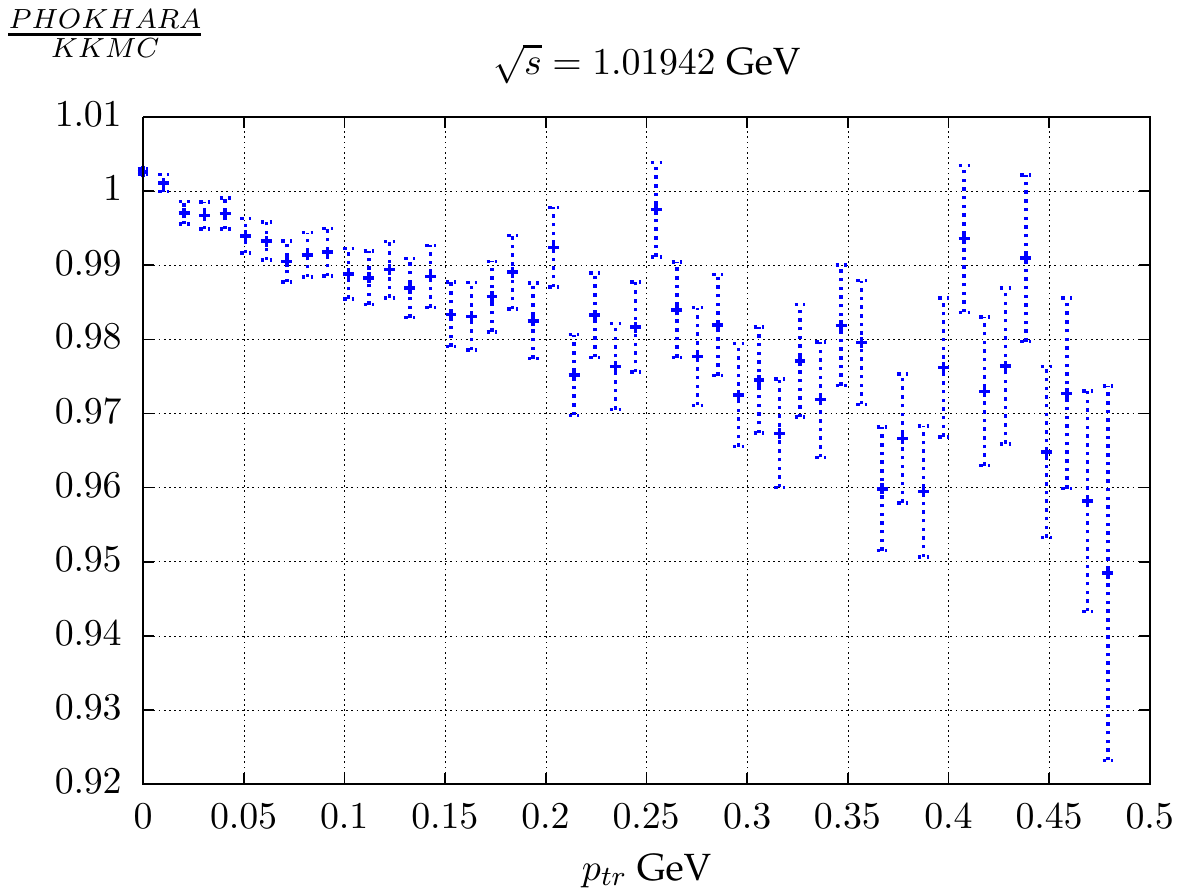}}
\end{center}
\vspace*{-8.0cm}
\caption[]{Comparison of the missing momentum distribution between PHOKHARA 8.0
 and
 KKMC 4.13 for $\sqrt{s} = 1.01942$~GeV, which are indistinguishable
in the left plot. }
\label{ptdistr}
\end{figure}

\begin{figure}[ht]
\begin{center}
\vspace*{-4.0cm}
\hspace*{-2.0cm}\subfloat{\includegraphics[width=.8\textwidth]{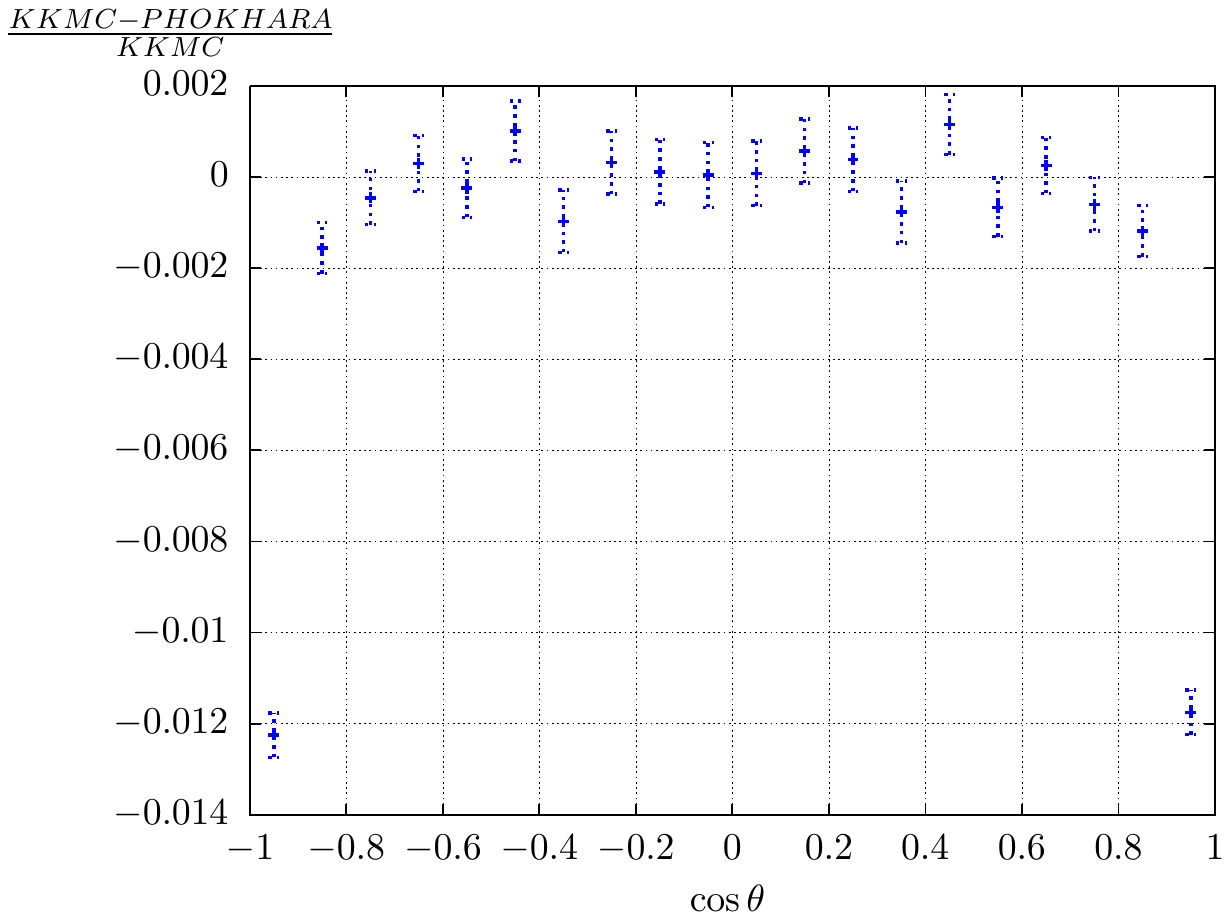}}
\hspace*{-4.0cm}\subfloat{\includegraphics[width=.8\textwidth]{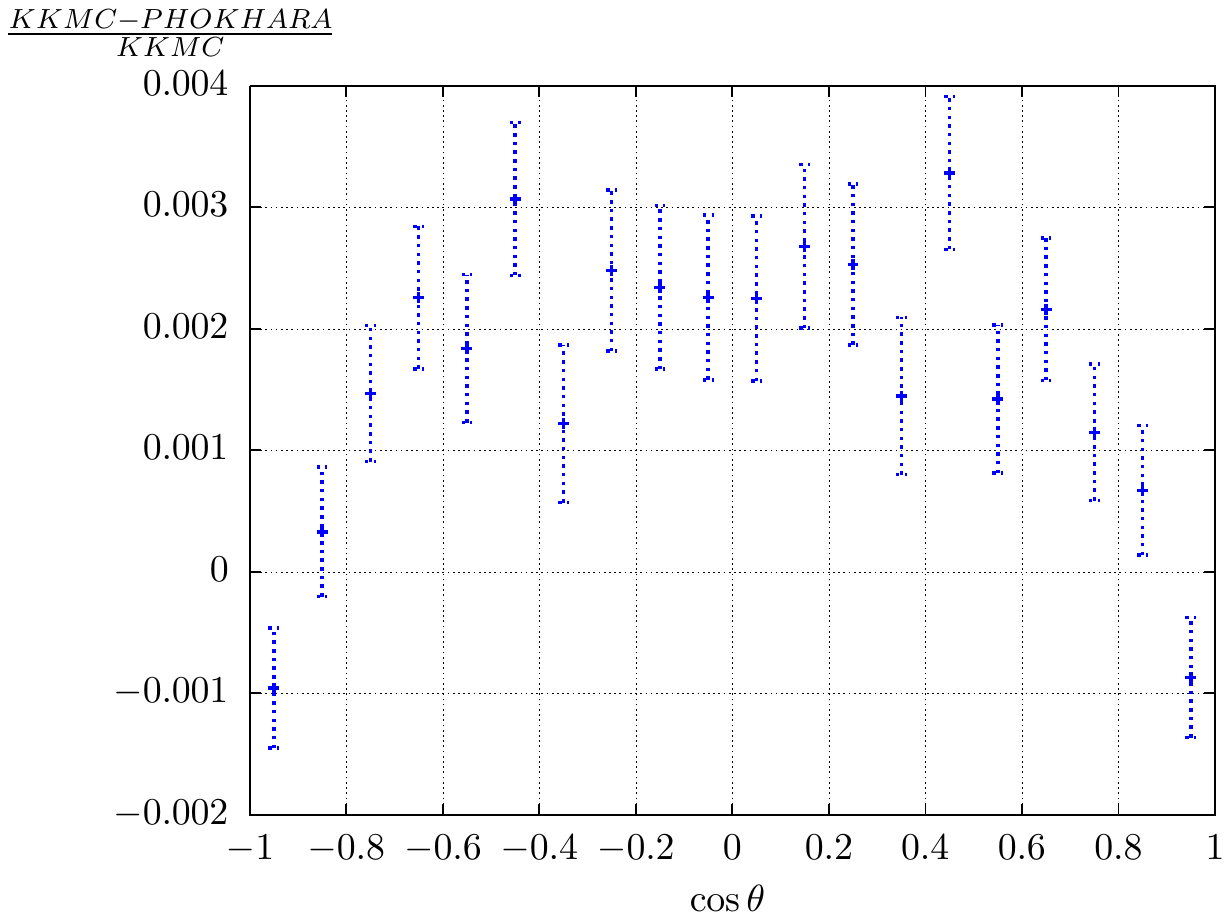}}
\end{center}
\vspace*{-8.0cm}
\caption[]{ The relative difference of the angular distributions
 given by PHOKHARA 8.0 and KKMC 4.13. In the left plot no cuts are applied.
 In the right plot the events with the muon pair invariant mass 
  $Q^2>$ 0.3~GeV are selected. }
\label{angmuons}
\end{figure}

  As another cross check of our new generator 
 the differential cross section with respect to the missing transverse momentum
 (carried away by one or two photons)  generated with
  PHOKHARA~8.0 and with KKMC 4.13 is shown in Fig. \ref{ptdistr}.
 The missing momentum is defined as the difference between
  incoming and outgoing fermion momenta. The small difference 
 between these distributions can be attributed
  to the missing multi-photon corrections in PHOKHARA  8.0 generator. 
A comparison of the angular distributions given by  KKMC 4.13 and PHOKHARA 8.0
codes is shown in Fig. \ref{angmuons}. The substantial difference 
 between the two codes seen in the left plot comes only from the region
 of invariant masses of the muon pairs close to the threshold and drops
 to permille level when the threshold region is excluded (right plot
 of Fig. \ref{angmuons}). We attribute this difference to the missing
 mass terms in the version of KKMC 4.13 we use and expect they will disappear
 when the missing mass corrections are included as in \cite{Jadach:2005gx}.

From the studies and comparisons presented in this section one can conclude
that the accuracy of the code, as far as ISR corrections are concerned,
 is from 0.3 permille for cumulative observables up to 2-3 percent
 for some specific differential distributions in regions of small event rates.

\section{Conclusions\label{sec-concl}}
%\addcontentsline{toc}{section}{Conclusions}
In its original form up to version 7.0 the generator PHOKHARA was  
constructed to simulate 
events with at least one photon from initial state radiation. 
In combination with the input based on the two-loop form factor 
\cite{Berends:1987ab} 
derived long time ago, the complete NNLO corrections from ISR 
are available and can be used to construct the corresponding generator,
now version 8.0,
for electron-positron annihilation into muon pairs or hadrons.
This generator is complementary to the two other Monte Carlo computer codes
currently in use, KKMC and MCGPJ. In particular PHOKHARA 8.0 does not include
multi-photon (beyond two) emission, however, it can be used to simulate
a multitude of exclusive hadronic final states and, in contrast to
structure function methods, the present approach includes exact
kinematics in the one- and two photon emission. Furthermore already in
its present version it includes a large number of exclusive hadronic final
states.

In the paper we demonstrate the independence of the results from
the soft photon cutoff and, for muon-pairs, compare a few selected
distributions with the results from other programs. Furthermore we study
the impact of NNLO compared to NLO corrections on the predicted cross
section and its dependence on the cut on the mass of the hadronic system
and find a strong dependence on the choice of the final state and 
 the center-of-mass energy of the experiment. 

Since our choice for the amplitude for the production of the three-meson 
state $\eta\pi^+\pi^-$ has not yet been documented in the literature, a
brief description is presented in the appendix.

\acknowledgments{
 We would like to thank Staszek Jadach for clarification of topics 
 concerning KKMC-PHOKHARA comparisons in \cite{Jadach:2005gx}. 
  We would like to thank our experimental friends for a constant
  push towards finishing this project, especially Achim Denig.
 H.C. would like to thank Simon Eidelman for discussions concerning 
  the $\eta \pi^+ \pi^-$ hadronic current.
  This work is a part of the activity of the ``Working Group on Radiative
 Corrections and Monte Carlo Generators for Low Energies" 
 [\href{http://www.lnf.infn.it/wg/sighad/}{http://www.lnf.infn.it/wg/sighad/}].
 This work is supported by BMBF project 05H12VKE,
 SFB/TR-9 ``Computational Particle Physics''
   and Polish Ministry of Science and High Education
   from budget for science for 2010-2013 under grant number N N202 102638.
  M. Gunia was supported by \'Swider PhD program.
   }

%%%%%%%%%%%%%%%%%%
%%%%%%%%%%%%%%%%%%
%%%%%%%%%%%%%%%%%%

\appendix

\section{Appendix: $\eta \pi^+ \pi^-$ hadronic current\label{app-eta}}
%\addcontentsline{toc}{section}{Appendix}

The differential cross section of the process  $e^+e^- \to \eta \pi^+ \pi^-$ can be written as:

\begin{equation}
d\sigma = \frac{1}{2s}\ |M|^2\ d\Phi(p_{e^+},p_{e^-};p_{\eta},p_{\pi^+},p_{\pi^-}),
\end{equation}
where
\begin{equation}
|M|^2 = L^{\mu\nu}H_{\mu\nu},\ \ H_{\mu\nu}=J_{\mu}J_{\nu}^*,
\end{equation}
\begin{equation}
L^{\mu\nu}=\frac{(4\pi\alpha)^2}{s^2}(p_{e^+}^{\mu}p_{e^-}^{\nu}+p_{e^-}^{\mu}p_{e^+}^{\nu}-g^{\mu\nu}\frac{s}{2}),
\end{equation}
and ${\rm d}\Phi(p_{e^+},p_{e^-};p_{\eta},p_{\pi^+},p_{\pi^-})$ denotes the three-body phase space.

 Isospin symmetry and charge-conjugation invariance restrict the
 $\eta\pi^+\pi^-$ system produced in $e^+e^-$ annihilation to isospin
 one. 

The hadronic current $J^\mu_{\rm em}$ responsible for this reaction can be
related to the corresponding charged current $J^\mu_{\rm cc}$ governing the
$\tau$ decay $\tau^+ \to \bar\nu \pi^+\pi^0$ \cite{Gilman:1987my}

\begin{equation}
J^\mu_{\rm cc} = \sqrt{2} \cos\theta_{\rm Cabibbo} J^\mu_{\rm em}
\end{equation}
Following \cite{Decker:1992kj,Jadach:1993hs} the amplitude is normalised 
 to its chiral limit, which is predicted by the Wess-Zumino-Witten term
 \cite{Wess:1971yu,Kramer:1984cy}. Its behaviour away from this point
 is governed by a form factor $F$, which includes the dominant
 $\rho$ resonance and its radial excitations and is normalised to one
 for $s=\left(p_\eta +p_{\pi^-}+p_{\pi^+}\right)^2=0$ and 
 $s_1= \left(p_{\pi^-}+p_{\pi^+}\right)^2=0$.
This leads to the following ansatz
\begin{equation}
J^{\mu}_{\rm em} = -\frac{i}{4\sqrt{3}\pi^2 f_{\pi}^3}\varepsilon^{\mu}_{\nu\rho\sigma}p_{\pi^+}^{\nu}p_{\pi^-}^{\rho}p_{\eta}^{\sigma} F,
\label{hadr_curr}
\end{equation}
with
%\begin{eqnarray}
%F &=&\frac{1}{1 + e^{i\phi} c_2 + e^{i\phi_0} c_0  } \ A_{\rho(770)}(s_1)\\ \nonumber
%\kern-20pt & \times &\left( A_{\rho(1450)}(s) + A_{\rho(1700)}(s) e^{i\phi} c_2 + A_{\rho(770)}(s) e^{i\phi_0} c_0 \right)
%\label{formfactorold}
%\end{eqnarray}

\begin{eqnarray}
F &=&\frac{1}{1 + e^{i\phi_1} c_1 + e^{i\phi_2} c_2  } \ BW_{\rho_0}(s_1)\\ \nonumber
\kern-20pt & \times &\left( BW_{\rho_0}(s) + BW_{\rho_1}(s) e^{i\phi_1} c_1 
 + BW_{\rho_2}(s) e^{i\phi_2} c_2 \right)
\label{formfactor}
\end{eqnarray}

\noindent
where $s_1$ is the $\pi^+\ \pi^-$ system invariant mass,

\begin{equation}
BW_{\rho_i}(s) = \frac{m_{\rho_i}^2
 }{m_{\rho_i}^2- s-i\sqrt{s}\Gamma_{\rho_i}(s)}
 \ ,
\end{equation}

\begin{equation}
\Gamma_{\rho_i}(s) = \Gamma_{\rho_i} \frac{s}{m_{\rho_i}^2},\ \ {\rm
 for} \ \ i=1,2
\end{equation}
and
\begin{equation}
\Gamma_{\rho_0}(s) = \Gamma_{\rho_0}
  \frac {{m_{\rho_0}^2}}{s}
 \cdot \left( \frac{s - 4 m_{\pi}^2}{m_{\rho_0}^2
- 4 m_{\pi}^2} \right)^{3/2},
\end{equation}
\noindent
$m_{\rho_0}$ = 0.77549 GeV and $\Gamma_{\rho_0}$ = 0.1494 GeV (PDG \cite{Amsler:2008zzb} values).

%\begin{table}[ht]
%\vspace*{1.0cm}
%\begin{center}
%\begin{tabular}{c|c|c|c}
%\hline
%$m_{\rho_1} $ & 1.470(7) [GeV] & $\Gamma_{\rho_1} $ &  0.28(2) [GeV] \\
%\hline
%$m_{\rho_2} $ & 1.76(7)[GeV] & $\Gamma_{\rho_2} $ & 0.3(1) [GeV] \\
%\hline
%$\phi_0 $ & 5.6(6) & $c_0 $ & -2.6(9) \\
%\hline
%$\phi $ & 1.7(9) & $c_2 $ & 0.2(1) \\
%\hline \hline
%$\chi^2$& 56 &$n_{d.o.f}$& 51 \\
%\hline
%\end{tabular}
%\caption{fit parameters}
%\label{fit_par}
%\end{center}
%\end{table}
%
%  
%
\begin{table}[ht]
\vspace*{1.0cm}
\begin{center}
\begin{tabular}{c|c|c|c}
\hline
$m_{\rho_1} $ & 1.470(7) [GeV] & $\Gamma_{\rho_1} $ &  0.28(2) [GeV] \\
\hline
$m_{\rho_2} $ & 1.76(7)[GeV] & $\Gamma_{\rho_2} $ & 0.3(1) [GeV] \\
\hline
$\phi_1 $ & -5.6(6) & $c_1 $ & -0.385(13) \\
\hline
$\phi_2 $ & -3.9(11) & $c_2 $ & -0.08(5) \\
\hline \hline
$\chi^2$& 56 &$n_{d.o.f}$& 51 \\
\hline
\end{tabular}
\caption{fit parameters}
\label{fit_par_new}
\end{center}
\end{table}

\begin{figure}
\vspace*{-5.0cm}
\begin{center}
\includegraphics[width=1.\textwidth]{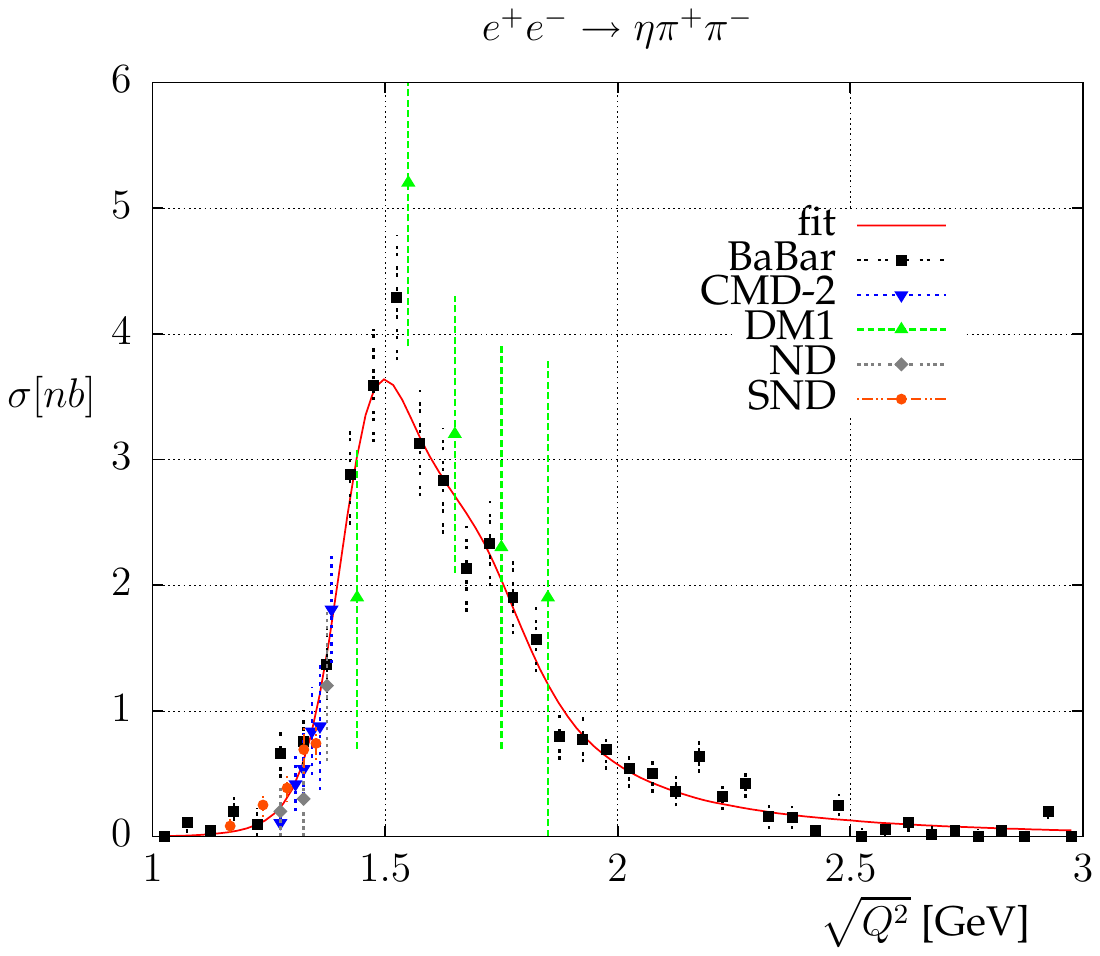}
\end{center}
\vspace*{-10.0cm}
\caption{Fitted cross section of $e^+e^-\to\eta\pi^+\pi^-$ and all experimental data used in the fit procedure: 
\cite{Aubert:2007ef,Akhmetshin:2000wv,Delcourt:1982sj,Antonelli:1988fw,Druzhinin:1986dk}.}
\label{fit}
\end{figure}

We performed an 8 parameter fit of the form factor (\ref{formfactor}) of the hadronic current (\ref{hadr_curr}) 
to the experimental data from BaBar \cite{Aubert:2007ef}, CMD-2 \cite{Akhmetshin:2000wv}, DM1 \cite{Delcourt:1982sj}, 
DM2 \cite{Antonelli:1988fw}, ND \cite{Druzhinin:1986dk} and SND.
Table \ref{fit_par_new} contains values of the fit's parameters together with its $\chi^2$, the result of the fit is compared to the data in Fig. \ref{fit}.

%\clearpage

\providecommand{\href}[2]{#2}
\addcontentsline{toc}{section}{References}
%%\bibliography{2loops}%,pub_lumi}
\bibliography{Ph0Ph}
%\bibliography{2loops}

\end{document}